\documentclass[showpacs,aps,prl,twocolumn,floatfix]{revtex4}
\usepackage{graphicx}

\begin{document}
\title{Asymptotically self-similar propagation of the spherical ionization waves}
\author{A. S. Kyuregyan}\email[E-mail: ]{ask@vei.ru}
\affiliation{All-Russian Electrical Engineering Institute, 12 Krasnokazarmennaya St., 111250 Moscow, Russian Federation}

\begin{abstract}
It is shown that a new type of the self-similar spherical ionization waves may exist in gases. All spatial scales and the propagation velocity of such waves increase exponentially in time. Conditions for existence of these waves are established, their structure is described and approximate analytical relationships between the principal parameters are obtained. It is notable that spherical ionization waves can serve as the simplest, but structurally complete and physically transparent model of streamer in homogeneous electric field.
\end{abstract}

\pacs{51.50.+v, 52.35.-g, 52.80.-s, 72.20.Ht}
\maketitle

The impulse electrical breakdown of various media is usually caused by appearance and propagation of the  ionization waves (IW), in particular, the streamers. However, the theory of this important and interesting subject of long-term investigations is still far from the completion. The reason for this state of affairs is quite obvious: it is necessary to solve multidimensional and nonlinear system of balance equations together with Poisson's equation for description of IW even in the framework of simplest hydrodynamic approximation. So the analytical description is possible only if they are highly symmetrical objects that possess the property of self-similarity. Among them stationary plane IW are the simplest ones and some rigorous results have been obtained for them~\cite{bt1,bt2,bt3}. Meanwhile the most of IW, studied experimentally and by numerical simulation, were neither stationary nor plane. Therefore subsequent search for the self-similar solutions, describing non-stationary curved IW, is of extreme importance. In this field first results were obtained recently: the authors of papers~\cite{bt4} studied cylindrical and spherical IW in gases, assuming the absence of full current and, therefore, the constancy of the wave charge value $Q$. This condition causes the reduction of the maximum field strength  $E_{M}\equiv|E(r_{f})|$  at the front with rise of its radius  $r_{f}$  in conformity with the same power law as for the initial undisturbed field. Such IW slow down and attenuate in the course of time and their evolution becomes self-similar only at so small  $E_{M}$,  that the impact ionization becomes negligible and the wave propagates only due to the electron drift.

In the present work I will examine a modified problem by assuming, in contrast to \cite{bt4}, that the current flows through the wave and supports the constancy of the surface charge density  ${Q/4\pi r_{f}^2}$  of the spherical wave front and, hence,  $E_{M}$. It will be shown, that in this case there exist asymptotically self-similar solutions, which describe undamped and exponentially accelerating IW. Such waves can serve the simplest, but structurally complete and physically transparent model of streamers in homogeneous electric field, for which the constancy of  $E_{M}$  is the most plausible hypothesis~\cite{bt5}.

Let us consider IW which propagates inside the spherical gas-filled capacitor towards the external electrode (for the certainty - to the anode). In the framework of ``minimal'' hydrodynamic model~\cite{bt2,bt4} the distributions of electrons  $n(t,r)$  and positively charged ions  $p\,(t,r)$  within the wave are described by the system of balance equations
\begin{equation}\label{eq1}
\frac{\partial n}{\partial t}+\nabla_{r}\left(vn-D\nabla_{r}n\right)=\frac{\partial p}{\partial t}=\alpha vn,
\end{equation}
\belowdisplayskip=0.6\belowdisplayskip
where the operator  $\nabla_{r}=r^{-2}\frac{\partial}{\partial r}r^{2}$. The impact ionization coefficient  $\alpha$, drift velocity  $v$  and diffusion coefficient  $D$  of electrons are considered as assigned instantaneous and local functions of the field strength  $E$. Usage of the first of equalities (\ref{eq1}) and Poisson's equation give the conservation law of the full current  $J\left(t\right)$  along the axis  $r$
\begin{equation}\label{eq2}
J\left(t\right)=4\pi qr^{2}\left(\frac{\varepsilon}{q}\frac{\partial E}{\partial t}-vn+D\nabla_{r}n\right),
\end{equation}
where $q$ is the elementary charge, $\varepsilon$ is the dielectric constant of the medium. Our goal is the search for solutions of the form  $f_{i}\left(\chi \right)\varphi_{i}\left(t\right)$, where self-similar coordinate  $\chi ={\psi \left(t/t_{0}\right)r/r_{0}}$, $t_{0}$ and $r_{0}$ are the positive constants and index $i=n,p,E$. It is possible to show, that allowing for impact ionization with any plausible dependence  $\alpha \left(E\right)$  this task is solvable only if $\psi \left(t/t_{0}\right)=\exp\left(-{t/t_{0}}\right)$  and every  $\varphi_{i}\left(t\right)=1$. In this case all addends in equations (\ref{eq1}),~(\ref{eq2}), that contain operator  $\nabla_{r}\propto\exp\left(-{t/t_{0}}\right)\nabla_{\chi}$, decrease exponentially in the course of time and at  $t\gg t_{0}$  our system takes the asymptotically self-similar form
\abovedisplayskip=0.7\abovedisplayskip
\begin{eqnarray}
\frac{d\sigma}{d\chi}=-\frac{\tau}{\chi}\sigma\left|V\right|A,\label{eq3}\qquad\\
\frac{dF}{d\chi}=\frac{\tau}{\chi}\left(\sigma V-\chi^{-2}\right).\label{eq4}
\end{eqnarray}
Here the following dimensionless variables and parameters are introduced: time  $\theta =\tilde{\alpha}\tilde{v}t$, concentration $\sigma\left(\chi \right)={qn/\varepsilon\tilde{\alpha}\tilde{E}}$, field strength  $F\left(\chi\right)={-E/\tilde{E}}$, drift velocity  $V\left(F\right)={v\left(E\right)/\tilde{v}}$, impact ionization coefficient  $A\left(F\right)={\alpha\left(E\right)/\tilde{\alpha}}$,   $\tau =\tilde{\alpha}\tilde{v}t_{0}$,  $\tilde{v}=v(\tilde{E})$,   $r_{0}=r_{c}({j_{0}/\varepsilon\tilde{\alpha }\tilde{v}\tilde{E}})^{1/2}$, where  $\tilde{\alpha}$  and  $\tilde{E}$  are constants, which characterize the dependence  $\alpha \left(E\right)$  (see further),  $r_{c}$  is the cathode radius,  $j_{0}$  is the initial current density at the cathode. During the derivation of equations (\ref{eq3}),~(\ref{eq4}) it has been assumed that  $J(t)=-4\pi r_{c}^{2}j_{0}\exp (2t/t_{0})$, since condition  $\varphi_{E}(t)=1$  is satisfied ahead of the front only if waves charge (and hence cathode current) grows according to the law  $Q\propto\exp(2t/t_{0})$. The number of equations was reduced by one in comparison with input system, since the space charge density  $q(p-n)\propto \exp(-t/t_{0})\nabla_{\chi}F|_{t\to\infty}\to 0$  and therefore  $n\approx p$  for all  $\chi $. Physical reason for this lies in the fact that at large  $\,t\,$  the front speed  $u_{f} =r_{f}/t_{0}\propto\exp(t/t_{0})$  becomes much greater than average electrons drift velocity. For the same reason equations (\ref{eq3}),~(\ref{eq4}) turn to be invariant with regard to simultaneous sign reversal of  $E$  and $j_{0}$. Consequently the parameters of self-similar IW do not depend on the electric field direction~\cite{bt5}.

The right sides of equations (\ref{eq3}),~(\ref{eq4}) and their second derivatives are continuous in terms of the variables  $\chi,\sigma,F$  at  $\chi>0$  and  $F>0$. Consequently solution of these equations exists and it is uniquely determined by boundary conditions ~\cite{bt6}. I will suppose that the radius of external electrode (anode)  $r_{a}\gg r_{f}$  and that there exist background charge carriers with the concentrations  $n_{+}=p_{+}$  ahead of the front. In this case the boundary conditions can be written as
\begin{equation}\label{eq5}
\sigma (\infty)=\sigma_{+}, \qquad F(\infty)=0.
\end{equation}

The complete solution of the boundary-value problem (\ref{eq3})-(\ref{eq5}) can be obtained only by numerical methods. To attain this goal I applied the Runge-Kut method of the 4th order with the adaptive grid, using the independent variable  $y={1/\chi}$  and solving the problem in the finite segment  $0<y<1/\chi_{c}$. The right boundary of this segment  $1/\chi_{c}=(j_{0}/\varepsilon\tilde\alpha\tilde{v}\tilde{E})^{1/2}\exp(\theta/\tau)$  corresponds to value  $r=r_{c}$ , i.e., the surface of internal electrode. Calculations were carried out using the simplest empirical approximations  $v=\mu\left|E\right|$ ($\mu$ is the electrons mobility) and $\alpha=\tilde\alpha\exp(-\tilde{E}/|E|)$, for which  $V(F)=F$  and  $A(F)=\exp(-1/F)$. The use of any other physically justified dependencies  $v(E)$  and  $\alpha(E)$  gives only quantitative changes of results.

\begin{figure}
\includegraphics[width=245pt]{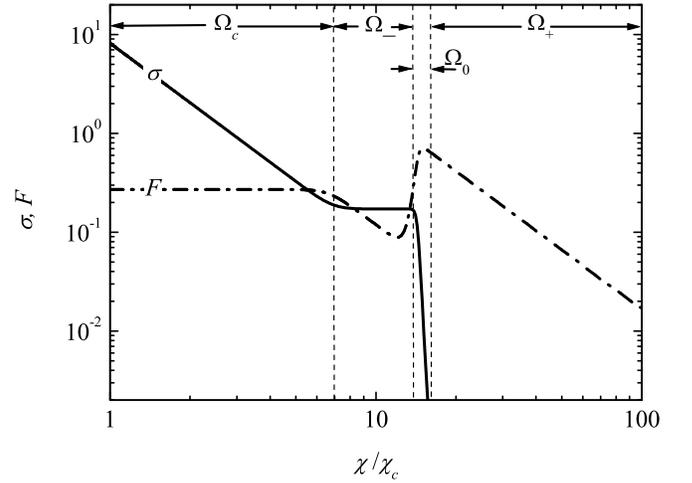}\\
\caption{Structure of spherical ionization wave at the moment  $\theta=5\tau$  with  $j_{0}=5\cdot 10^{-5}\varepsilon \tilde{\alpha}\tilde{E}\tilde{v}$,  $\tau =300$  and  $\sigma_{+}=10^{-6}$.}\label{f1}
\end{figure}

Typical distributions of field and concentration in spherical IW are presented in Fig.~\ref{f1}. Results for other instants of time can be obtained by a simple shift along the logarithmic  $\chi$-axis. It is evident, that the wave consists of four qualitatively different regions  $\Omega_{c}$,  $\Omega_{-}$,  $\Omega_{0}$  and  $\Omega_{+}$. In the region  $\Omega_{+}$  ahead of the front the conductivity is negligible, therefore it follows from~(\ref{eq4}) that
\begin{equation}\label{eq6}
F\approx\frac{\tau}{2\chi^{2}}
\end{equation}
This formula describes field distribution around the charged sphere. Substituting~(\ref{eq6}) in~(\ref{eq3}) one can get after integration the interrelation between concentration and field strength
\begin{equation}\label{eq7}
\sigma\approx\sigma _{+}\exp\left\{\frac{\tau}{2}[\eta_{0}-\eta(F)]\right\},
\end{equation}
where $\eta \left( F \right) = \int_F^{F_M } {AdF} $ and $\eta _0 = \eta\left( 0 \right)$. These formulae are valid until $\sigma\tau\ll 2$. Fast growth of conductivity at the wave front leads to the appearance of the region $\Omega _0$, in which the field first reaches maximum value $F_{M}$ with $\chi=\chi_{M}$ and then begin decrease. In this region  both suppression of field due to the plasma conductivity and abrupt change of this conductivity due to intense impact ionization are important factors. The thickness $\delta$ of region $\Omega_0$ usually much less then $\chi_{M}$ and the field strength changes insignificantly in $\Omega _0$. Therefore it is possible to assume $\chi=\chi_{M}$ and $V(F)=F_M$ in equations (\ref{eq3}),(\ref{eq4}), but sharp exponential dependence $A(F)$ should be allowed. For this case the solution of (\ref{eq3}),(\ref{eq4}) was obtained in paper \cite{bt3} and with $V(F)=F$ it take the form
\abovedisplayskip=1.5\abovedisplayskip
\begin{eqnarray}
\chi\approx\chi_M\left({1-\frac{1}{\tau F_M}\int\limits_F^{F_M }{\frac{dF}{\eta -\sigma _M\lambda}}}\right),\label{eq8}\\
\sigma\approx\sigma _M\exp(-\lambda), \qquad \qquad \label{eq9}
\end{eqnarray}
where $\sigma _M=\sigma\left({\chi_M}\right)$, and $\lambda$ is positive (with $\chi>\chi_{M})$ or negative (with $\chi<\chi_{M})$ root of the equation
\begin{equation}\label{eq10}
\lambda-1+\exp(-\lambda)=\eta/\sigma_M.
\end{equation}
Field reduction behind the wave front causes the formation of the region $\Omega_{-}$, where impact ionization is negligible and the solution takes the form
\begin{eqnarray}
F\approx\frac{F_{m}}{\tau\sigma_{-}+2}\left[2\left(\frac{\chi}{\chi_{m}}\right)^{\tau \sigma_{-}}+\tau\sigma_{-}\left(\frac{\chi_{m}}{\chi}\right)^{2}\right],\label{eq11}\\
\sigma\approx\sigma_{-}=\frac{1}{F_{m}\chi_{m}^{2}},\qquad \qquad \qquad\label{eq12}
\end{eqnarray}
where  $F_{m}=F\left(\chi_{m}\right)$  is minimal field in the wave. With  $\chi <\chi_{m}$  the field strength increases in the direction to the cathode, that is caused by the need for conservation of the full current through plasma  with constant conductivity  $\sigma=\sigma_{-}$  and with decreasing area  $4\pi r^{2}\propto\chi^{2}$. Since usually  $\tau\sigma_{-}\gg 2$, the field increases as  $F\approx F_{m}(\chi_{m}/\chi)^2$  until in the region  $\Omega_{c}$ ($\chi<\chi_{-}$) recurring growth of  $\sigma$  begins due to the impact ionization. As a result of drastic dependence  $A\left(F\right)$  sufficient growth of  $\sigma$  is ensured due to the  small increase of  $F$. Therefore near the cathode field strength approaches to constant value  $F_{c}$, while conductivity increases according to power law:
\abovedisplayskip=0.5\abovedisplayskip
\begin{equation}\label{eq13}
F\approx F_{c}, \qquad \sigma\approx\sigma_{c}(\chi)\equiv\frac{1}{F_{c}\chi^{2}} \label{eq12}.
\end{equation}
If  $F_{c}$  is the root of equation  $\tau F_{c}A(F_{c})=2$, then formulae (\ref{eq13}) give one of exact solutions of the equations~(\ref{eq3}),~(\ref{eq4}). This solution does not satisfy to boundary conditions~(\ref{eq5}), but  it is attractor for the real solutions with  $\chi\to 0$ (see Fig.~\ref{f1}). On the cathode itself the conductivity  $\sigma =1/(F_{c}\chi_{c}^{2})$  and increases exponentially with the increment  $2/t_{0}$  in the course of time, as it must be to provide the same increase of the current.

In order to obtain complete description of IW it is necessary to join the solutions (\ref{eq6})-(\ref{eq13}) between themselves. Taking into account equality $\sigma_M F_M\chi_M^2=1$ the formula (\ref{eq6}) can be rewritten in the form $\tau\sigma_M=2(\chi_{0M}/\chi_M)^2$, where $\chi_{0M}=\sqrt{\tau/2F_M}$  by definition. Smallness $\delta$
makes it possible to disregard a difference between $(\chi_{0M}/\chi_M)^2$ and 1. As a result the
estimations of concentration $\sigma _M\approx 2/\tau$, front position $\chi_f\equiv\chi_{M}\approx\sqrt{\tau/2F_{M}}$
and front speed $u_f=r_f/t_0=\left(r_0 {t_0}^{-1}\sqrt{\tau/2F_{M}}\right)\exp(t/t_0)$ are derived immediately. Joining dependencies (\ref{eq7}), (\ref{eq9}) and taking into account the fact that ahead of the front $\lambda\approx(\eta/\sigma_M+1)\approx(\tau\eta/2+1)$, it is easy to obtain approximate equation
\begin{equation}\label{eq14}
\ln\frac{2}{\tau\sigma_{+}}\approx\frac{\tau}{2}\eta_0+1,
\end{equation}
that establish the relation of maximum field $F_{M}$ with control the parameters $\sigma_{+}$, $\tau$. Formula for the electron concentration behind the front $\sigma_{-}$ follows from (\ref{eq9}):
\abovedisplayskip=1.5\abovedisplayskip
\begin{equation}\label{eq15}
\sigma_{-}=\frac{2}{\tau}\exp\left({-\lambda_{-}}\right),
\end{equation}
where $\lambda_-$ is the negative root of equation $\left[{\lambda_{-}-1+ \exp\left({-\lambda_{-}}\right)}\right]= \eta_0\tau/2$. Instead of (\ref{eq15}) it is possible to use the simple approximation
\begin{equation}\label{eq16}
\sigma_{-}=a_\sigma\eta_0,\quad a_\sigma=1+\frac{9}{2\ln{\eta_0/\sigma _{+}}}
\end{equation}
that differs from (\ref{eq15}) less than to 2{\%} under condition $\eta_0>15\sigma_{+}$, which is satisfied well usually. The analysis of the numerical solutions of the boundary-value problem (\ref{eq3})-(\ref{eq5}) shows, that  $\left(\chi_{0M}/\chi_M\right)^2<1.1$ with $\sigma_{+}\le 10^{-4}$. This inequality justifies the approximations $\chi_{0M}=\chi_{M}$ used above. The stright comparison with numerical results shows that  formulae (\ref{eq14})-(\ref{eq16}) are carried out with a good accuracy in all examined cases, as it is evident from Fig. 2,3.    Further, equalizing conductivity currents on the surfaces $\chi=\chi_{m}$ and $\chi=\chi_{M}$, one can obtain relationship for estimation of $F_{m}$:
\abovedisplayskip=0.7\abovedisplayskip
\begin{equation}\label{eq17}
\sigma_{-}F_m=2a_\chi F_M/\tau,
\end{equation}
where $a_\chi=\left(\chi_{0M}/\chi_m\right)^2$. Formula (\ref{eq17}) ensures high accuracy if $a_\chi=1.15a_\sigma$ (see Fig.4). Finally, join of solutions (\ref{eq11})-(\ref{eq13}) give the position of boundary $\chi_{-}\approx(\sigma_{-}F_{c})^{-1/2}$ between regions  $\Omega_{c}$ and  $\Omega_{-}$.

\begin{figure}[tbp]
\includegraphics[width=245pt]{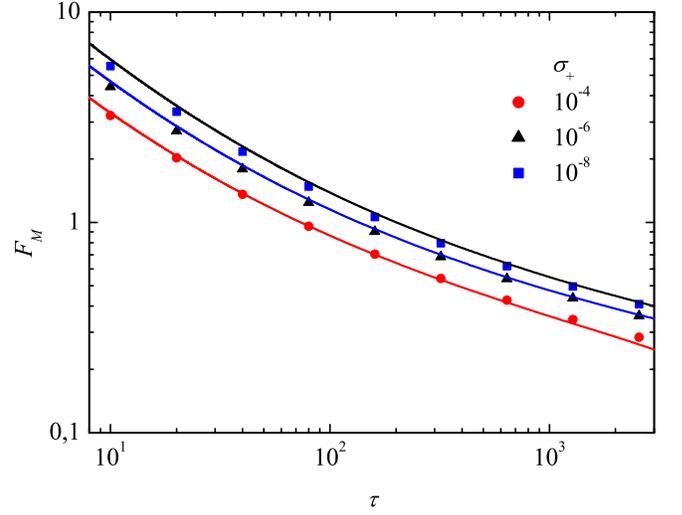}\\
\caption{Plot of  maximum field strength  against response time  with different values of  $\sigma_{+}$. Points: numerical solution, line: calculation according to (\ref{eq14}).}\label{f2}
\end{figure}

\begin{figure}[tbp]
\includegraphics[width=245pt]{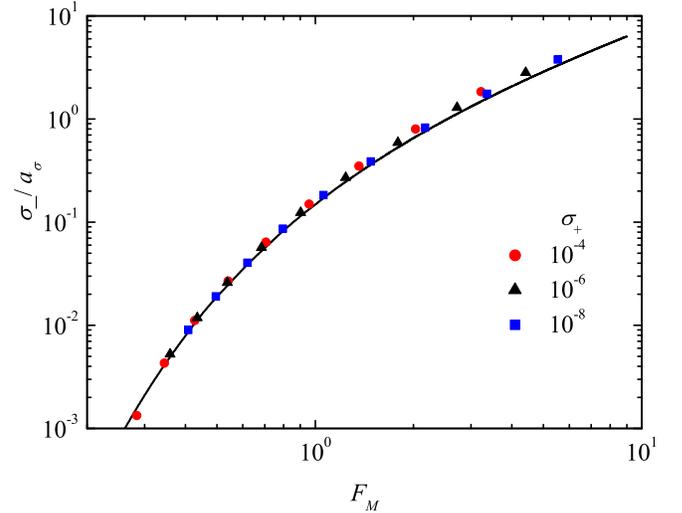}\\
\caption{Plot of  electron concentration   behind the front against maximum field strength  with different values of  $\sigma_{+}$. Points: numerical solution, lines: calculation according to (\ref{eq16}).}\label{f3}
\end{figure}

The obtained formulae establish interrelations of all principal values, that define the structure of IW ($F_{c}$,  $\chi_{-}$, $\chi_{m}$,  $\sigma_{-}$, $F_{m}$, $\chi_{f}$, $F_{M}$), with the control parameters ($\sigma_{+}$,  $\tau $). Thus they give the complete description of waves for large  $t$. It is clear, however, that the acceleration of IW cannot continue infinitely for a long time. Firstly, waves must be impeded due to the appearance of vortex electrical field, when the front velocity approaches to the velocity of light \cite{bt5}. Secondly, exponential build-up of plasma conductivity near the cathode sooner or later will be suppressed by nonlinear recombination, electron-ion scattering, saturation of cathode emission ability or due to the total ionization of medium. The above mentioned and similar mechanisms will fail to appear, if a radius  $r_{a}$  of external electrode is not too great. In this case self-similar propagation of the wave will continue up to the front contact with the external electrode at the time  $t\approx t_{0}\ln (r_{a}/r_{0}\chi_{f})$, if external electric circuit ensures the exponential build-up of current. It is very difficult to attain this in practice since value  $\tau=1$  corresponds to response time  $t_{0}$  of the order of several picoseconds. Therefore a reasonable question arises concerning the existence of the real objects, for which a model of self-similar accelerating IW is descriptive.

\begin{figure}[tbp]
\includegraphics[width=245pt]{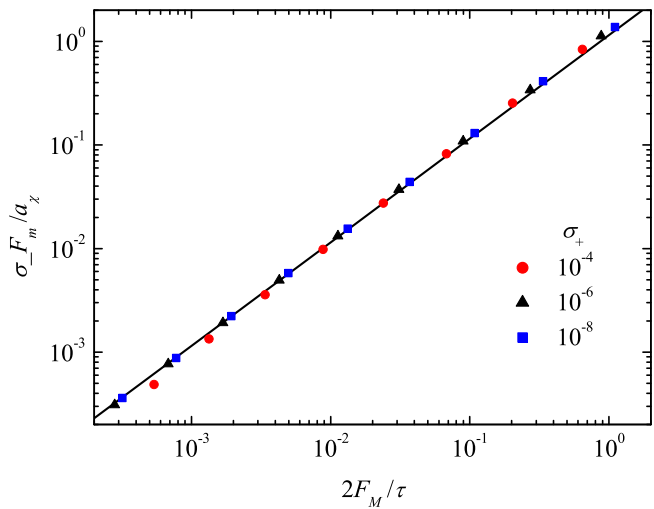}\\
\caption{Plot of conductivity current density  behind the front against displacement current density ahead of the front  with different values of  $\sigma_{+}$. Points: numerical solution, line: calculation according to (\ref{eq17}) with  $a_{\chi}=1.15a_{\sigma}$}\label{f4}
\end{figure}

In our opinion such objects are, in particular, the streamers in uniform electric field, which repeatedly were studied by two-dimensional numerical simulation. The most detailed data are represented in the papers \cite{bt7} for nitrogen at normal pressure. Analysis of these data indicates the qualitative coincidence of structure and evolution characteristics of streamer and spherical self-similar IW. This coincidence is so complete that (being distracted from the nicety of the streamers transverse structure) it is possible to declare: streamer in uniform electric field is simply a cone with spatial angle at vertex of order  $0.1\pi$, that have been ``cut out'' from self-similar accelerating spherical IW. However, the evolution of such streamers inside a large parallel-plate capacitor possesses a number of features in comparison with spherical IW. First of all, an exponential growth of the charge with the increment  $2/t_{0}$ (which is necessary for the self-similar propagation) is ensured not due to external control, but automatically during evolution of streamer as the self-organizing object, that obtains energy from the uniform field. Secondly, in gases the seeding electrons ahead of the streamers front arises mainly due to ionizing radiation, emitted from plasma behind the front. In this case IW can propagate almost self-similarly as long as  $r_{f}\alpha_{ph}<1$, where  $\alpha_{ph}$  is minimal absorption coefficient of the ionizing radiation. This inequality was fulfilled in \cite{bt7}, where  a value  $\alpha_{ph}=5{\rm\;cm}^{{\rm-1}}$ was used in calculations up to the value $r_{f}\sim 0.2{\rm\;cm}$. Thirdly, streamers evolution must lead to an increase in the current in external circuit exponentially with the increment  $3/t_{0}$, although the increment of charge growth is equal to  $2/t_{0}$ just as in the spherical wave. This nontrivial result was revealed, but not explained in \cite{bt7}. It follows directly from Sato theorem \cite{bt8} and from the condition for exponential growth of all streamers spatial scales with the increment  $1/t_{0}$.

In conclusion it should be noted that self-similar IW can also exist even if uniform distributed seeding charge carriers are absent (i.e., at  $\sigma_{+}=0$), but some additional mechanism of medium ionization exists ahead of the front. The effective action radius $r_{g}$  of this mechanism, obviously, must increase in proportion to $r_{f}$. This scenario can be realized in semiconductors, where tunnel ionization is the basic mechanism of electron-hole pairs generation ahead of the front \cite{bt9}. The rate of this process depends on $E$  locally. Hence condition $r_{g}\propto r_{f}$  is satisfied automatically and streamer in the uniform field must  propagate in accordance with the described scenario for a reasonably long time.

I thank A. V. Gorbatyuk for many helpful discussions.

\end{document}